\documentclass[iop]{emulateapj} 
\usepackage{natbib}
\usepackage{epstopdf}
\newcommand{\hiir}{H~{\scshape ii}~region}
\newcommand{\hi}{H~{\scshape i}}
\newcommand{\hiirs}{H~{\scshape ii}~regions}

\newcommand{\degree}{^{\circ}}
\newcommand{\msol}{M$_{\odot}$}

\newcommand{\methanol}{CH$_3$OH}
\slugcomment{Submitted to ApJ Letters, 21 June 2013}

\shorttitle{}
\shortauthors{}

\begin{document}

%% LaTeX will automatically break titles if they run longer than
%% one line. However, you may use \\ to force a line break if
%% you desire.

\title{Early stage massive star formation near the Galactic Center: Sgr C}

%% Use \author, \affil, and the \and command to format
%% author and affiliation information.
%% Note that \email has replaced the old \authoremail command
%% from AASTeX v4.0. You can use \email to mark an email address
%% anywhere in the paper, not just in the front matter.
%% As in the title, use \\ to force line breaks.

\author{S. Kendrew\altaffilmark{1}}
\affil{Max-Planck-Institut f\"{u}r Astronomie, K\"{o}nigstuhl 17, 69117 Heidelberg, Germany}
\email{kendrew@mpia.de}
\altaffiltext{1}{University of Oxford, Department of Astrophysics, Keble Road, Oxford OX1 3RH, United Kingdom}

\author{A. Ginsburg}
\affil{CASA, University of Colorado at Boulder, UCB 389, Boulder, CO 80309, USA}

\author{K. Johnston}
\affil{Max-Planck-Institut f\"{u}r Astronomie, K\"{o}nigstuhl 17, 69117 Heidelberg, Germany}

\author{H. Beuther}
\affil{Max-Planck-Institut f\"{u}r Astronomie, K\"{o}nigstuhl 17, 69117 Heidelberg, Germany}  

\author{J. Bally}
\affil{CASA, University of Colorado at Boulder, UCB 389, Boulder, CO 80309, USA} 

\author{C.J. Cyganowski\altaffilmark{2}}
\affil{Harvard-Smithsonian Center for Astrophysics, Cambridge, MA 02138, USA}
\altaffiltext{2}{NSF Astronomy and Astrophysics Postdoctoral Fellow}

\author{C. Battersby}
\affil{CASA, University of Colorado at Boulder, UCB 389, Boulder, CO 80309, USA}

%\author{M.G. Burton}
%\affil{School of Physics, University of New South Wales, Sydney, NSW 2052, Australia}

%% Notice that each of these authors has alternate affiliations, which
%% are identified by the \altaffilmark after each name.  Specify alternate
%% affiliation information with \altaffiltext, with one command per each
%% affiliation.

%\altaffiltext{1}{Visiting Astronomer, Cerro Tololo Inter-American Observatory.
%CTIO is operated by AURA, Inc.\ under contract to the National Science
%Foundation.}
%\altaffiltext{2}{Society of Fellows, Harvard University.}
%\altaffiltext{3}{present address: Center for Astrophysics,
%    60 Garden Street, Cambridge, MA 02138}
%\altaffiltext{4}{Visiting Programmer, Space Telescope Science Institute}
%\altaffiltext{5}{Patron, Alonso's Bar and Grill}

%% Mark off your abstract in the ``abstract'' environment. In the manuscript
%% style, abstract will output a Received/Accepted line after the
%% title and affiliation information. No date will appear since the author
%% does not have this information. The dates will be filled in by the
%% editorial office after submission.

\begin{abstract}  
We present near-infrared spectroscopy and 1~mm line and continuum observations of a recently identified site of high mass star formation likely to be located in the Central Molecular Zone near Sgr C. Located on the outskirts of the massive evolved~\hiir~associated with Sgr C, the area is characterized by an Extended Green Object measuring $\sim$10\arcsec~in size (0.4 pc), whose observational characteristics suggest the presence of an embedded massive protostar driving an outflow. Our data confirm that early-stage star formation is taking place on the periphery of the Sgr C~\hiir, with detections of two protostellar cores and several knots of H$_2$ and Brackett $\gamma$ emission alongside a previously detected compact radio source. We calculate the cores' joint mass to be $\sim$10$^3$~\msol, with column densities of 1-2$\times$10$^{24}$ cm$^{-2}$. We show the host molecular cloud to hold $\sim$10$^5$~\msol~of gas and dust with temperatures and column densities favourable for massive star formation to occur, however, there is no evidence of star formation outside of the EGO, indicating that the cloud is predominantly quiescent.  Given its mass, density, and temperature, the cloud is comparable to other remarkable non-star-forming clouds such as G0.253 in the Eastern CMZ.
\end{abstract}

%% Keywords should appear after the \end{abstract} command. The uncommented
%% example has been keyed in ApJ style. See the instructions to authors
%% for the journal to which you are submitting your paper to determine
%% what keyword punctuation is appropriate.


%% From the front matter, we move on to the body of the paper.
%% In the first two sections, notice the use of the natbib \citep
%% and \citet commands to identify citations.  The citations are
%% tied to the reference list via symbolic KEYs. The KEY corresponds
%% to the KEY in the \bibitem in the reference list below. We have
%% chosen the first three characters of the first author's name plus
%% the last two numeral of the year of publication as our KEY for
%% each reference.

%% Authors who wish to have the most important objects in their paper
%% linked in the electronic edition to a data center may do so by tagging
%% their objects with \objectname{} or \object{}.  Each macro takes the
%% object name as its required argument. The optional, square-bracket 
%% argument should be used in cases where the data center identification
%% differs from what is to be printed in the paper.  The text appearing 
%% in curly braces is what will appear in print in the published paper. 
%% If the object name is recognized by the data centers, it will be linked
%% in the electronic edition to the object data available at the data centers  
%%
%% Note that for sources with brackets in their names, e.g. [WEG2004] 14h-090,
%% the brackets must be escaped with backslashes when used in the first
%% square-bracket argument, for instance, \object[\[WEG2004\] 14h-090]{90}).
%%  Otherwise, LaTeX will issue an error. 

\section{Introduction}\label{sec:intro}

The Central Molecular Zone (CMZ) of the Milky Way Galaxy is a chemically and dynamically complex region containing up to 10\% of the Galaxy's molecular gas, concentrated in dense and turbulent molecular clouds ($-2.5\degree\leq l \leq 3.5\degree$, $|b| \leq 0.5 \degree$). Densities and turbulent velocities are known to be approximately an order of magnitude higher than clouds in the MW disk~\citep{Morris1996}, and shocks are widely observed~\citep{Riquelme2010}. The CMZ presents an ideal laboratory for the study of star formation in extreme environments.

%the commonly accepted star formation relations that are widely used as templates for studies of star formation in galaxies throughout cosmic time. 

The distribution of molecular gas in the CMZ is asymmetric around the Galactic Center (GC), with roughly two-thirds of the gas found at positive (Eastern) longitudes; this is mirrored in the observed distribution of CMZ star formation \citep[YZ09 hereafter]{Yusef-Zadeh2009a}. The Eastern CMZ is home to several star-forming regions, e.g. Sgr B2.

\begin{figure*} [h]
 %   \centering
	\includegraphics[width=18cm]{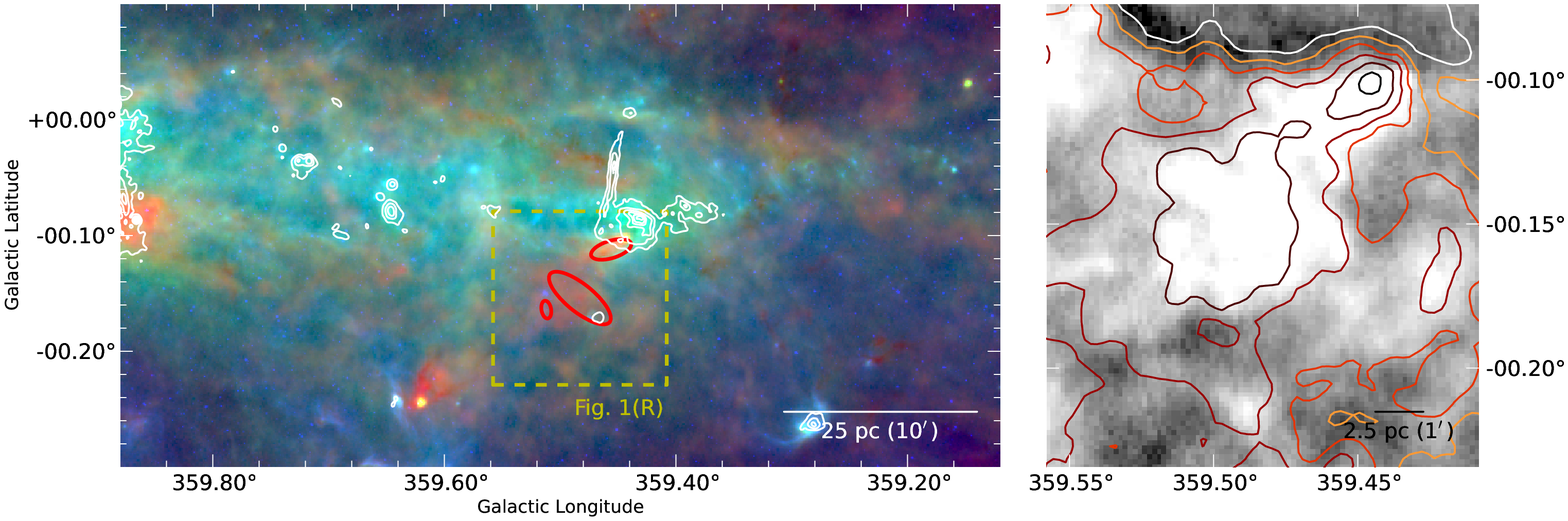}
	\caption{Left: Wide-field view of Sgr C (105 $\times$ 75 pc) showing 8/70/870~\micron~data from the GALCEN~\citep[blue]{Stolovy2006}, Hi-GAL~\citep[green]{Molinari2010a} and ATLASGAL surveys~\citep[red]{Schuller2009}, respectively. VLA 1.4 GHz continuum data from~\citet{Lang2010} are shown in the white contours (levels: 20\%, 40\%, 60\%, 80\%, 90\%, 100\% of the peak flux of 0.208 Jy beam$^{-1}$). Red ellipses show the ATLASGAL clumps from the~\citet{Contreras2012} catalog. Right: Grayscale 870~\micron~ATLASGAL image with Hi-GAL-derived temperature contours. Darker contours indicate lower temperatures (levels: 20, 22, 24, 26, 28 and 30K).}\label{fig:widefield}
\end{figure*}

The only known star forming region in the Western CMZ is Sgr C (Figure~\ref{fig:widefield}). \citet{Lang2010} identify the region's main components as a 10-pc~\hiir~at $(l, b) =$(359.43, -0.09) (assuming a distance of 8.5 kpc;~\citealt{Reid2009}), accompanied by a distinctive non-thermal filament~\citep{liszt95}. The~\hiir~contains more than 250 \msol~of ionized gas, powered by at least one O4-O6 star~\citep{liszt95, Odenwald1984}, however its stellar population remains poorly characterized. To the East of the \hiir~lies a  pillar-shaped cloud, whose mass has been estimated at 10$^5$~\msol~\citep{Lis1991, Lis1994}.   

\citet{Molinari2011} recently proposed that Sgr C lies at the Western vertex of a twisted elliptical ring of molecular gas and dust, 100 $\times$ 60 pc in size. The Eastern longitude extremum is formed by Sgr B2. They propose that the collision of orbit systems at the extrema give rise to strong shocks, driving enhanced star formation. Given the prolific star formation rate of Sgr B2, Sgr C must be examined more closely to understand the observed asymmetry in the CMZ.

The first evidence of high mass star formation in the Sgr C cloud was reported by~\citet{Forster2000}, who detected a faint 8-9 GHz radio source measuring 0.06 $\times$ 0.01 pc near its tip. Using data from the Spitzer GALCEN survey~\citep{Stolovy2006}, YZ09 identify a region of extended 4.5~\micron~emission - a so-called Extended Green Object (EGO;~\citealt{Cyganowski2008}) - within $\sim$5\arcsec~from the radio source. Numerous studies have found EGOs to be strongly associated with early-stage high mass star formation and outflows~\citep{DeBuizer2010, Cyganowski2009}. Three~\methanol~masers are seen within the EGO at velocities consistent with the Sgr C systemic velocity (-55 to -65 km s$^{-1}$), supporting this scenario~\citep{Caswell2009, Chambers2011}. In this Letter, we present new  infrared and millimeter data towards the EGO, which we refer to as G359.44-0.102, showing the presence of two massive protostellar cores and emission knots indicative of an outflow.

\section{Data and observations}\label{sec:data}
 
\emph{TripleSpec near-infrared spectroscopy.} To identify the origin of G359.44-0.102's infrared emission, we carried out medium-resolution (R$\sim$100 km s$^{-1}$) spectroscopy in the near-infrared K-band, using TripleSpec on the 3.5-m ARC Telescope at Apache Point Observatory, in August 2012 (PI: Ginsburg). The K-band contains several H$_2$ emission lines, most notably the 1-0 S(1) line at 2.12~\micron, frequently observed in outflows from YSOs \citep{Varricatt2010, Davis2010}. Spectra were obtained at 4 positions with a 1.1\arcsec~slit. The data were taken in ABBA slit-nodding mode with a series of 30s exposures, for a total integration time of 2-6 minutes per pointing.  

\emph{Submillimeter Array (SMA).} High-resolution data to search for dense star-forming gas toward the G359.44-0.102 EGO were obtained in June 2009 with the compact-north configuration of SMA at 280 GHz (PI: Kauffmann). These data consist of a mosaic of five $\sim$45\arcsec~fields covering the EGO and its surrounding clump with spectral resolution of 0.44 km s$^{-1}$. The observations covered the N$_2$H$^+$ (3-2) line at 279.512 GHz, commonly used to trace dense gas, with RMS noise of 0.23 Jy beam$^{-1}$ when imaged at 2 km s$^{-1}$ resolution. A combined self-calibrated continuum image was produced from the 2 GHz sidebands centred at 278.8 and 288.8 GHz, with combined RMS noise of 0.012 mJy beam$^{-1}$. The synthesized beam of the continuum and line images is $\sim$2\arcsec. Data reduction and analysis were performed using CASA.

\emph{Survey data.} The CMZ was covered by a number of Galactic Plane surveys. The Herschel Hi-GAL survey \citep{Molinari2010a} was used to derive temperature estimates toward the EGO and its host cloud, as described in \citet{Battersby2011}. A Galactic cirrus emission model was estimated using an improved iterative technique which varies the source threshold value until it converges, classifying everything below that threshold as part of the diffuse cirrus. The emission model at each wavelength was subtracted from each of the five Hi-GAL bands, which were fit with a modified blackbody assuming $\beta$ = 1.75 to derive the dust temperature at each point (Figure~\ref{fig:widefield}; $\beta$ derived from \citet{osshenning94} opacity law). The 870~\micron~ATLASGAL survey~\citep{Schuller2009} traces cold dense gas at a spatial resolution of $\sim$19\arcsec (Figure~\ref{fig:widefield}). Fluxes at this wavelength were used to determine the physical properties of the EGO's host cloud~\citep{Contreras2012}. Spectral line data at 3-mm from the Mopra CMZ survey~\citep{Jones2012} were used to determine the velocity of the EGO's region. Data on masers in the region were obtained from~YZ09,~\citet{Chambers2011} and~\citet{Caswell2009}.

\section{Results}\label{sec:results}    

\subsection{TripleSpec near-infrared spectroscopy}

\begin{figure*}[h]
	\centering
	\includegraphics[width=15cm]{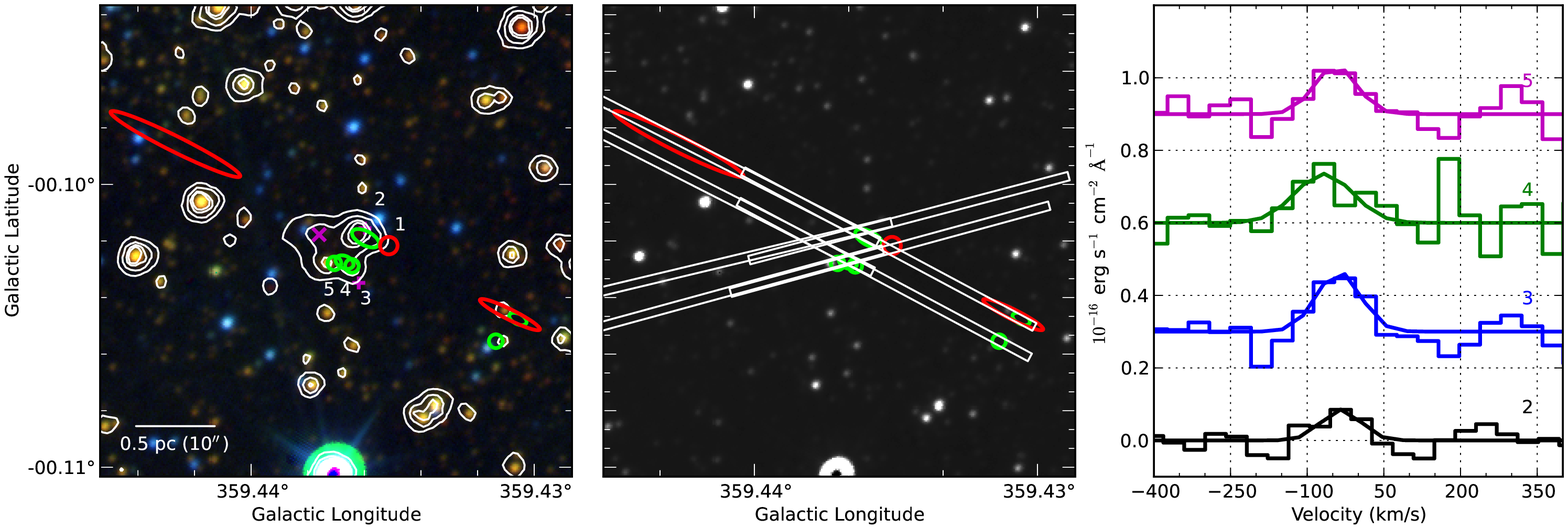}
	\caption{Left: J/H/K band image of G359.44-0.102. White contours show 4.5~\micron~emission, which marks the extent of the EGO (levels: 50, 100, 200, 300 MJy/sr).~\methanol~masers are shown in magenta, 6.7 GHz as +, 44 GHz as x~\citep{Caswell2010}. Green markers show the H$_2$ 2.12~\micron~emission knots from the TripleSpec spectra; red markers show the Br~$\gamma$ emission. Numbering of the knots matches Table~\ref{tab:tspec}. Middle: Greyscale image showing the TripleSpec slit positions. Right: Spectra of the H$_2$ lines.}\label{fig:nir}
\end{figure*}

The near-infrared spectra reveal a number of emission locations towards the EGO and in its vicinity (Figure~\ref{fig:nir}). Within the extent of the 4.5~\micron~emission we find 4 knots of H$_2$ emission at 2.12~\micron~in 2 clusters approximately 5\arcsec~(0.2 pc) apart. The Eastern cluster contains three knots, with Gaussian-fitted v$_{LSR}$ of -30, -50 and -70 km s$^{-1}$; the Western knot has a velocity of -40 km s$^{-1}$. The lines are unresolved at the TripleSpec spectral resolution. 

Near the Western H$_2$ knot we find a site of Br~$\gamma$ emission at -25 km s$^{-1}$. Its proximity to the radio source (1\arcsec) appears consistent with the presence of a hyper- or ultra-compact~\hiir; the velocity discrepancy may be the result of dynamical interaction between UC\hiir~and the host cloud.

Several emission features are detected along the slits away from the EGO. Their association with the region is unclear, and we do not include these further in our analysis. Properties of the detected emission features are summarised in Table~\ref{tab:tspec}. As the slit positions do not cover the extent of the IR emission, the detected knots may extend beyond the slit boundaries. Further knots may be found given better coverage.

\begin{table*}
	\centering
	\begin{tabular}{cccccc}
		\hline
		Index & $\alpha$ (J2000) & $\delta$ (J2000) & Line & V (km s$^{-1}$) & Line flux ($\times$ 10$^{-16}$ erg s$^{-1}$ cm$^{-2}$)\\
		\hline
		%1 & 17:44:40.058 & -29:28:35.65 & H$_2$ & +28 $\pm$ 3 & \\
		%2 & 17:44:40.064 & -29:28:34.49 & Br~$\gamma$ & +20 & \\
		1 & 17:44:40.107 & -29:28:16.77 & Br~$\gamma$ & -25 $\pm$ 4  & 19.12 $\pm$ 0.02\\
		2 & 17:44:40.163 & -29:28:13.75 & H$_2$ & -34 $\pm$ 12 & 6.41 $\pm$ 0.02\\
		%5 & 17:44:40.348 & -29:27:46.88 & Br~$\gamma$ & 0 & \\
		%6 & 17:44:40.355 & -29:28:34.73 & H$_2$ & +21 $\pm$ 7 & \\
		3 & 17:44:40.462 & -29:28:14.12 & H$_2$ & -39 $\pm$ 9 & 11.85 $\pm$ 0.02\\
		4 & 17:44:40.469 & -29:28:13.07 & H$_2$ & -66 $\pm$ 27 & 13.77 $\pm$ 0.04 \\
		5 & 17:44:40.532 & -29:28:12.05 & H$_2$ & -47 $\pm$ 15 & 9.11 $\pm$ 0.03\\
		\hline
	\end{tabular}
	\caption{Emission features detected in the near-infrared ARC/TripleSpec spectra, numbered as in Figure~\ref{fig:nir}. }\label{tab:tspec}
\end{table*}

\subsection{SMA 280 GHz data}

\begin{figure*}
	\centering
	\includegraphics[width=15cm]{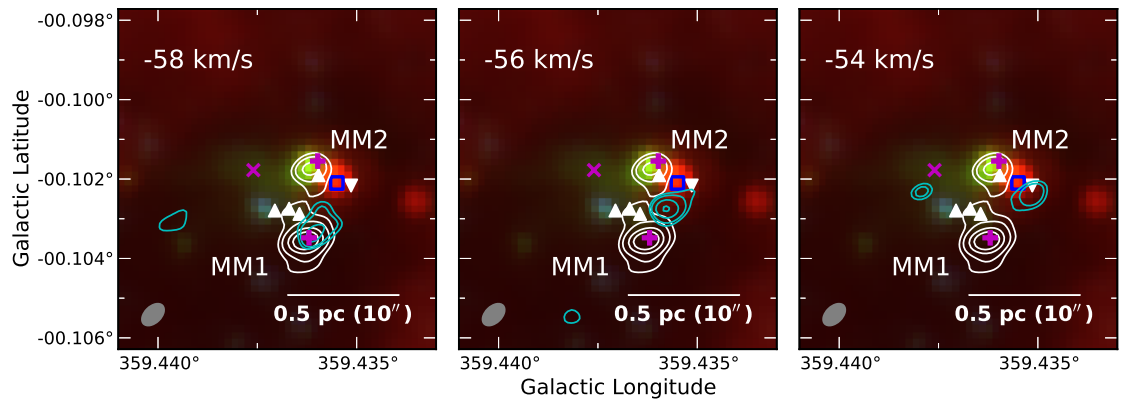}
	\caption{3.6/4.5/8.0~\micron~image of the EGO with SMA 280 GHz contours. White contours represent continuum emission from combined LSB/USB (levels: 5, 10, 20 and 30 $\sigma$ of 12 mJy~beam$^{-1}$). Cyan contours show spectral line emission from N$_2$H$^+$ (3-2) at 279.512 GHz (levels: 5, 6, 8, 10 $\sigma$ of 0.23 Jy beam$^{-1}$). The maps cover a velocity width of 2 km~s$^{-1}$, centred at -58, -56 and -54 km~s$^{-1}$ from left to right. White triangles and upside-down triangles mark the locations of the H$_2$ and Br~$\gamma$ emission knots. Blue squares show the location of the compact radio source from~\citet{Forster2000}. Magenta symbols are as in Figure~\ref{fig:nir}.}\label{fig:sma}
\end{figure*}

The 280~GHz SMA data show two sources in the continuum bands associated with the EGO. Figure~\ref{fig:sma} shows the location of these sources with respect to the infrared emission region. 

The brightest, MM1, is located $\sim$6\arcsec~(0.25 pc) South from the peak of the 4.5~\micron~emission, coincident with a 6.7~GHz maser~\citep{Caswell2010}. It measures 5.5 $\times$ 5\arcsec~in size (0.2 $\times$ 0.2 pc), measured to the 5-$\sigma$ contour, and is elongated in the North-South direction (different from the beam elongation direction). At the location of the IR emission peak lies a second fainter source, MM2, also associated with a 6.7~GHz~\methanol~maser. Measuring 3 $\times$ 3\arcsec~in size (0.12 $\times$ 0.12 pc), MM2 shows a similar elongation along the N-S axis. Neither core appears associated with the radio source, suggesting the presence of at least 3 distinct star-forming sites. The integrated continuum fluxes measured within the 1-$\sigma$ contours are 1073~mJy for MM1 and 623~mJy for MM2. We assumed uncertainties on these values of 15\%, dominated by the flux calibration accuracy.

In addition, the SMA data show N$_2$H$^+$ line emission near the brighter core, with v$_{LSR}$ ranging from -58 to -54 km~s$^{-1}$. Channel maps and continuum data are shown in Figure~\ref{fig:sma}.

\section{Discussion}\label{sec:discussion}

\subsection{Distance to G359.44-0.102}\label{disc:distance}
 
Dynamics of Sgr C are complex, with evidence of strong velocity gradients~\citep{liszt95} and numerous absorption components in \hi~spectra~\citep{Lang2010}. The velocity of the near 3 kpc arm at a distance of 5.5 kpc is very similar to the Sgr C systemic value at $l$ = -0.5~\citep{Oka1998}, causing potential confusion; we note that \citet{Green2009} place the \methanol~masers in the 3 kpc arm.

\begin{figure*}[h]
    \includegraphics[width=13cm]{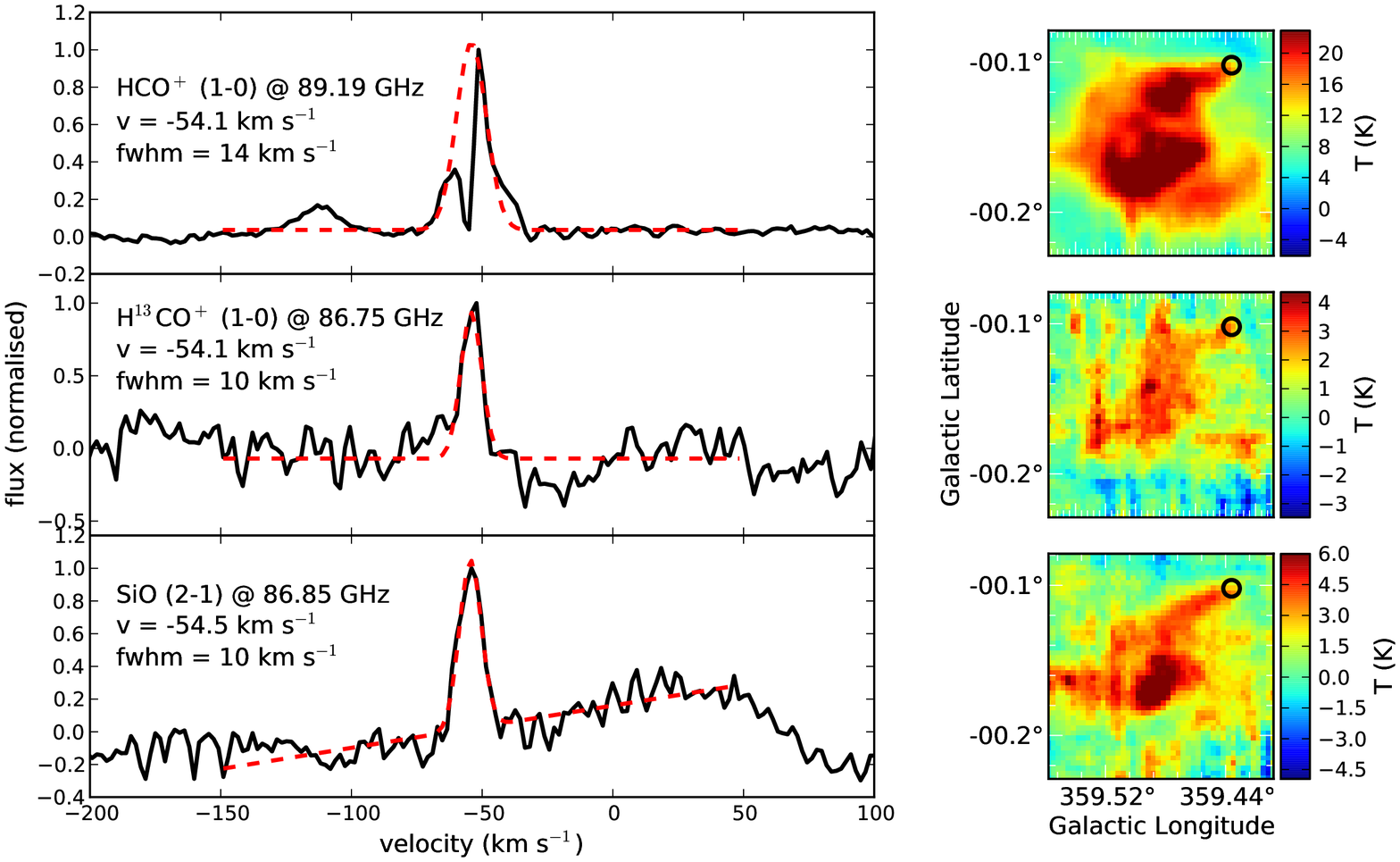}
    \caption{Mopra CMZ survey spectra (left), with (right) the corresponding integrated velocity map of the Sgr C cloud. The maps were integrated from -45 to -75 km~s$^{-1}$, and cover the region of Fig.~\ref{fig:widefield}(R). Spectra are extracted in a 40\arcsec~aperture centred on G359.44-0.102 (see black circles) and spatially averaged. Black lines show the data, red dotted lines are the best-fit Gaussians. Errors on fit parameters are $<$ 1 km s$^{-1}$}\label{fig:mopra}
\end{figure*}

3-mm spectra from the Mopra CMZ survey~\citep{Jones2012} were extracted in a 40\arcsec~aperture (the Mopra beam size) centred on the EGO for HCO$^+$ (1-0) at 89.19 GHz, H$^{13}$CO$^+$ (1-0) at 86.75 GHz and SiO (2-1) at 86.85 GHz (Figure~\ref{fig:mopra}; the poor baseline calibration in weaker species is discussed by~\citealt{Jones2012}). We show integrated velocity maps (-45 to -75 km~s$^{-1}$) alongside.

The spectra show strong emission peaks near -54 km~s$^{-1}$ with fitted peak velocities consistent to within 1 km~s$^{-1}$. Linewidth measurements indicate full widths at half maximum (FWHM) of $>$10 km~s$^{-1}$. Such broad linewidths are highly characteristic of CMZ clouds~\citep{Oka1998, Shetty2012}; observed velocities for disk clouds are $\leq$ 5 km~s$^{-1}$~\citep{Shetty2012}.

The SiO (2-1) integrated velocity map shows widespread emission throughout the cloud. Extended SiO emission is a well-documented feature of CMZ clouds~\citep{MartinPintado1997, Riquelme2010}; in the disk SiO emission is typically observed in outflow shocks locally.  

These observed characteristics suggest that the EGO and its host cloud are located at the GC distance. Some uncertainty does however remain, the masses calculated in the following sections would in this case be reduced by 40\%.

\subsection{Physical conditions in G359.44-0.102}

Integrated flux measurements from the SMA continuum data were used to estimate masses of the protostellar cores in the EGO using the following relation~\citep{hildebrand83}:

\begin{equation}
	M = \frac{g S_{\lambda} D^2}{\kappa_{\lambda}B_{\lambda}(T_d)}
\label{eqn:mass}\end{equation}

\noindent where $g$ is the gas-to-dust conversion ratio, $S_{\lambda}$ is the integrated flux at the observing wavelength $\lambda$, $D$ is the distance to the source, $\kappa_{\lambda}$ the dust mass opacity coefficient at $\lambda$, and $B_{\lambda}(T_d)$ is the blackbody flux for a dust temperature $T_d$ evaluated at $\lambda$. From the Hi-GAL-derived temperature map (Figure~\ref{fig:widefield}) we see that the EGO lies near the coldest part of the cloud at a temperature of $\sim$20K. We assume $g$ of 76~\citep[chap. 23]{Draine2011}, calculated using metallicity $Z_{GC} = 2 Z_{\odot}$~\citep{Launhardt2002}, and D of 8.5 kpc. 

For the dust opacity coefficient~$\kappa$ we log-interpolate the data in Table 1 of~\citet{osshenning94}. For a gas density of 10$^6$~cm$^{-3}$ and grains with thin ice mantles, we find $\kappa_{1.06}$ of 1.25~cm$^2$~g$^{-1}$ for the SMA frequency of 280 GHz.%; $\kappa_{0.87} = 1.75~cm^2~g^{-1}$ for ATLASGAL; and $\kappa_{1.1} = 1.18~cm^2~g^{-1}$ for BGPS. 

For the brighter continuum source MM1 the integrated flux yields a mass of 668~\msol. Source MM2 has a mass of 380~\msol. These values are averages of the masses calculated from the individual sideband fluxes. We estimate the combined uncertainty of these mass estimates to be a factor of 2-4, based on estimates on uncertainty in flux calibration, T$_d$, $\kappa$ and g. 

Similarly, peak column densities were computed using

\begin{equation}
	N(H_2)^{peak} = \frac{S_{\lambda}\; g}{B_{\lambda}(T_d)\; \Omega\; \kappa_{\lambda}\; \mu m_H}
\label{eq:coldens}\end{equation}

\noindent where $\Omega$ is the beam solid angle, $\mu$ is the mean molecular weight of the interstellar medium (assumed to be 2.8), $m_H$ the mass of a hydrogen atom, and $g$, $S_{\lambda}$ and $B_{\lambda}(T_d)$ as defined above. Using the peak measured fluxes and taking a 2\arcsec~SMA beam, we find peak column density estimates of 2 $\times$ 10$^{24}$ cm$^{-2}$ and 1.1 $\times$ 10$^{24}$ cm$^{-2}$ for sources MM1 and MM2, respectively.

\subsection{Evidence for star formation in G359.44-0.102}

The observational characteristics and derived properties of G359.44-0.102 are similar to those of other EGOs reported in the literature. In a sample of 28 objects, ~\citet{Cyganowski2009} found 64\% of EGOs to be associated with 6.7 GHz Class II~\methanol~maser emission, and of those 89\% additionally have 44 GHz Class I masers. Using mm and radio observations,~\citet{Cyganowski2011b,Cyganowski2011, Cyganowski2013} demonstrate that EGOs are reliable markers of massive star formation, often harbouring multiple star forming cores with strong active outflows. 

The SMA continuum data show the presence of two star forming cores, each carrying several 100~\msol~in mass with high column densities, alongside the UC\hiir~proposed by \citet{Forster2000} . There is no evidence of significant dust heating towards these sources, nor of free-free or Br~$\gamma$ emission suggesting a UC\hiir~has begun to form. The two main sites of H$_2$ emission are clustered near the mm cores (Figure~\ref{fig:nir}), with velocities both blue- and redshifted with respect to the systemic value.

In the absence of additional line detections, we cannot establish the excitation mechanism of these lines; however, the properties and locations of the knots show compelling evidence for the presence of at least one massive outflow. The detected mm cores are in close proximity ($\leq$ 0.5 pc) to the compact radio source, indicating that the EGO hosts at least three distinct sites of massive star formation.

\subsection{G359.44-0.102 in a quiescent host cloud}

The EGO is located at the tip of a dense molecular cloud seen in absorption against the bright IR background. Survey data from the ATLASGAL and Hi-GAL surveys give an insight into the nature of this cloud.

The cloud measures roughly 7 $\times$ 4\arcmin~in ATLASGAL maps, corresponding to $\sim$16 $\times$ 9 pc at the GC distance. The Hi-GAL temperature map suggests temperatures across the cloud lie in the range of 20-25 $\pm$ 2 K, with a minimum of 19 K coinciding with the peak of the sub-mm emission. 

Source extraction from ATLASGAL data by~\citet{Contreras2012} identify three sources within the cloud (AGAL359.437-00.102, AGAL359.474-00.152, AGAL359.514-00.154; Figure~\ref{fig:widefield}), measuring 3$\times$1 pc, 4$\times$2 pc and 1$\times$0.6 pc. \citet{Contreras2012} report integrated fluxes of 119 Jy, 168 Jy and 25 Jy. Assuming parameters for Equation~\ref{eqn:mass} as listed above and finding $\kappa_{0.87}$ of 1.75~cm$^2$~g$^{-1}$, these values yield gas masses of $\sim$4, $\sim$5 and $\sim$0.8$\times$10$^4$~\msol. The total mass enclosed in the dense cloud is thus of the order of 10$^{5}$~\msol, consistent with previous estimates~\citep{Lis1991, Lis1994}. The mass contained in cores MM1 and MM2 ($<$10$^3$~\msol) represents just a small fraction of this total mass.

The clump-averaged column densities of the ATLASGAL sources were calculated using the effective radii from the Contreras catalog, and found to be in the range of 1.5-3.5 $\times$ 10$^{22}$ cm$^{-2}$. These values indicate that the entire cloud shows favourable physical conditions for massive stars to form~\citep{Lada2012}. Apart from the data presented here for the EGO in source AGAL359.437-00.102, however, we found no 8, 24 or 70~\micron~sources in the cloud indicative of further star formation activity. No radio emission is observed and no~\methanol~masers are reported in the literature. The EGO's location at the cloud tip is likely to be pertinent, with star formation perhaps accelerated at the interface of a cloud-cloud collision or by feedback from the~\hiir. A detailed study of the dynamics of the region is required to understand which mechanism is at work.

The cloud shares similarities with the massive compact cloud G0.253+0.016~\citep{Lis1994b, Longmore2012, Kauffmann2013}. The latter has a similar mass (1-2 $\times$ 10$^5$~\msol) as the Sgr C cloud, albeit in an apparently smaller volume, yet shows no evidence of significant star formation activity. \citet{Immer2012} report the discovery of a further 4 compact 10$^5$~\msol~clouds in the CMZ that appear devoid of star formation. 

At the other extreme, the Sgr B2 star forming region is one of the most prolific and well-studied in the Galaxy. Using 1.1~mm maps from the BGPS, \citet{Bally2010} find a mass of 5 $\times$ 10$^5$~\msol~and average H$_2$ column densities of $>$10$^{24}$ cm$^{-2}$ towards its main star-forming sites. Its chemistry is however far richer, many of its star formation sites being more evolved hot cores and UC\hiirs. 

Based on our current data, Sgr C lies between the CMZ star-forming extrema represented by the G0.253 and Sgr B2. Despite the new evidence of star formation presented here, the Sgr C cloud as a whole still appears to be quiescent compared with non-CMZ clouds showing similar physical conditions~\citep{Lada2010}. This adds to the evidence presented by~\citet{Longmore2013} and~\citet{Kauffmann2013} for an anomalously low star formation rate in the CMZ. A more detailed study of Sgr C is clearly required to establish the region's global star formation properties.

\section{Conclusions}  

We have presented new, archival and survey data of the molecular cloud thought to be associated with Sgr C. Our high-resolution infrared and millimeter data show two massive protostellar cores associated with an EGO alongside a previously detected hyper- or ultra-compact~\hiir, which appear to be driving at least one outflow. The absence of evidence for dust heating indicates that these cores are at an early evolutionary stage. 

The cores are deeply embedded in the densest region of a $\sim$ 16 $\times$ 9 pc molecular cloud abutting the Sgr C~\hiir. We note that the measured velocities also allow this site to be associated with the 3 kpc arm at 5.5 kpc, however linewidths and widespread SiO emission indicate a likely distance of 8.5 kpc. The cloud harbours $\sim$10$^5$~\msol~of gas and dust with typical column densities of $>$10$^{22}$ cm$^{-2}$ and temperatures of 19-25 K. It shows no evidence of further star formation activity despite presenting suitable physical conditions. 

The presence of three star-forming cores in the EGO contrasts with the absence of further star formation in the host molecular cloud, placing Sgr C at an interesting mid-point between the quiescent G0.253 and starburst-like Sgr B2. The driving and inhibiting forces at work in the CMZ clouds and the relevance of their location along the proposed 100-pc elliptical ring are fascinating open questions.

\section{Acknowledgements}
\acknowledgements{

SK thanks Jens Kauffmann, Sergio Molinari, Cornelia Lang and Yanett Contreras for sharing and discussing data. The Apache Point Observatory 3.5-m telescope is owned and operated by the Astrophysical Research Consortium. The Submillimeter Array is a joint project between the Smithsonian Astrophysical Observatory and the Academia Sinica Institute of Astronomy and Astrophysics. The Mopra radio telescope is part of the ATNF, funded by the Commonwealth of Australia.  

}

\facility{
{\it Facilities:} \facility{Spitzer (IRAC, MIPS)}, \facility{ARC}, \facility{Mopra}, \facility{APEX}, \facility{UKIRT}
}

%% In a manner similar to \objectname authors can provide links to dataset
%% hosted at participating data centers via the \dataset{} command.  The
%% second curly bracket argument is printed in the text while the first
%% parentheses argument serves as the valid data set identifier.  Large
%% lists of data set are best provided in a table (see Table 3 for an example).
%% Valid data set identifiers should be obtained from the data center that
%% is currently hosting the data.
%%
%% Note that AASTeX interprets everything between the curly braces in the 
%% macro as regular text, so any special characters, e.g. "#" or "_," must be 
%% preceded by a backslash. Otherwise, you will get a LaTeX error when you 
%% compile your manuscript.  Special characters do not 
%% need to be escaped in the optional, square-bracket argument.

%% Putting eqnarrays or equations inside the mathletters environment groups
%% the enclosed equations by letter. For instance, the eqnarray below, instead
%% of being numbered, say, (4) and (5), would be numbered (4a) and (4b).
%% LaTeX the paper and look at the output to see the results.

\bibliographystyle{apj}
%\bibliography{sgrc_300413_new}  

\end{document}